\documentclass[aps,prl,reprint,superscriptaddress]{revtex4-1}

\usepackage[dvipdfmx]{graphicx}

\begin{document}


\title{Infrared activation of the Higgs mode by supercurrent injection in
superconducting NbN}


\author{Sachiko Nakamura}
\affiliation{Cryogenic Research Center, the University of Tokyo, Yayoi, Tokyo, 113-0032, Japan}
\author{Yudai Iida}
\affiliation{Department of Physics, the University of Tokyo, Hongo, Tokyo, 113-0033, Japan}
\author{Yuta Murotani}
\affiliation{Department of Physics, the University of Tokyo, Hongo, Tokyo, 113-0033, Japan}
\author{Ryusuke Matsunaga}
\affiliation{The Institute for Solid State Physics, the University of Tokyo, Kashiwa, Chiba 277-8581, Japan}
\author{Hirotaka Terai}
\affiliation{National Institute of Information and Communications Technology, 588-2 Iwaoka, Nishi-ku, Kobe 651-2492, Japan}
\author{Ryo Shimano}
\affiliation{Cryogenic Research Center, the University of Tokyo, Yayoi, Tokyo, 113-0032, Japan}
\affiliation{Department of Physics, the University of Tokyo, Hongo, Tokyo, 113-0033, Japan}

\date{\today}

\begin{abstract}
Higgs mode in superconductors, i.e. the collective amplitude mode of the order parameter does not associate with charge nor spin fluctuations, 
therefore it does not couple to the electromagnetic field in the linear response regime. 
On the contrary to this common understanding, here, we demonstrate that, if the dc supercurrent is introduced into the superconductor, the Higgs mode becomes infrared active and is directly observed in the linear optical conductivity measurement. 
We observed a sharp resonant peak at $\omega=2\Delta$ in the optical conductivity spectrum of a thin-film NbN in the presence of dc supercurrent, showing a reasonable agreement with the recent theoretical prediction. The method as proven by this work opens a new pathway to study the Higgs mode in a wide variety of superconductors.  
\end{abstract}


\maketitle

\begin{figure}[bp]
\includegraphics[width=.49\textwidth]{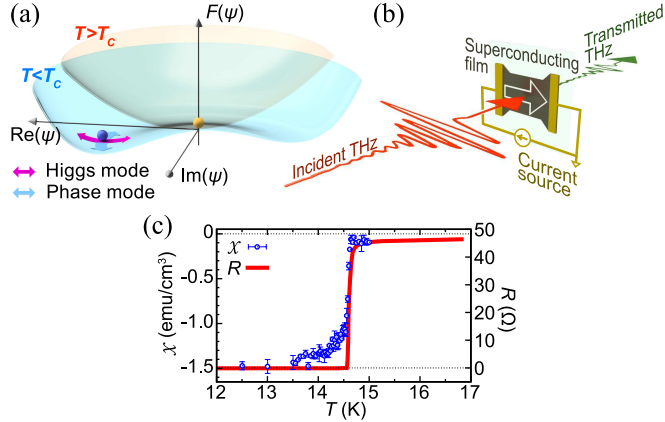}%
\caption{(a) Free energy $F(\Psi)$ with respect to the superconducting order parameter $\Psi$. The Higgs (amplitude) mode and the phase mode are represented by the arrows. (b) Schematic view of the terahertz transmittance experiments under the dc current. 
(c) Temperature dependence of the ac magnetic susceptibility $\chi$ measured at 1\,Oe in 1488\,Hz (open symbols) and electrical resistance $R$ (closed symbols).}\label{fig1}
\end{figure}

The collective modes ubiquitously exist in a variety of systems, e.g. in charge density wave (CDW) systems, ferromagnets and antiferromagnets, superfluid $^4$He and $^3$He, cold atomic gas systems, and superconductors, providing insights into the nature of symmetry broken ground states. 
In general, two types of collective modes emerge when a continuous symmtery is spontaneously broken: the phase mode and amplitude mode that correspond to the fluctuation of phase and amplitude of the order parameter, respectively (see Fig.~\ref{fig1}(a)). 
In superconductors, the phase mode is lifted up to the high energy plasma frequency because of the screening of long range Coulomb interaction~\cite{Anderson1963}. 
The remaining amplitude mode, recently referred to as the Higgs mode, has gained a growing interest over decades~\cite{Littlewood1982,Pekker2015}. 
Since the initial prediction made by Anderson~\cite{Anderson1963}, intensive theoretical studies have been devoted to elucidate the energy structure, stability, and relaxation mechanism of the Higgs mode to date. 
The behavior of Higgs mode has been discussed from the viewpoint of order parameter dynamics after the quantum quench~\cite{Volkov1973,Barankov2004,Yuzbashyan2005,Yuzbashyan2006,Barankov2006,Papenkort2007,Papenkort2008,Gurarie2009,Schnyder2011,Foster2013,Krull2014,Kemper2015,Peronaci2015,Murakami2016,Krull2016,Muller2018}, which has been recently observed by terahertz pump-probe experiments in s-wave superconductors~\cite{Matsunaga2013}. 
The coupling of the Higgs mode to the gauge field was initially identified in the Raman signal with the aid of strong electron-phonon coupling in NbSe$_2$, where the CDW coexists with superconductivity~\cite{Littlewood1982,Sooryakumar1980,Measson2014,Grasset2018}. 
Even in an s-wave superconductor without the CDW order, the nonlinear coupling between the Higgs mode and the gauge field has been elucidated in THz pump-probe response and third harmonic generation~\cite{Matsunaga2014, Matsunaga2017,Katsumi2018}, and extensive microscopic theories have been developed to date~\cite{Tsuji2016,Jujo2015,Cea2016,Behrle2018,Yu2017}. 
The observability of Higgs mode in the linear response, i.e. in the optical conductivity has been addressed in two-dimensional disordered superconductors~\cite{Podolsky2011}. 
Experimentally, the finite spectral weight below the superconducting gap 2$\Delta$ observed in a disordered ultrathin NbN film sample was attributed to Higgs mode from the comparison with tunneling spectroscopy~\cite{Sherman2015}, whereas different origins of the spectral weight have also been suggested, i.e. disorder-induced broadning of the quasiparticle density of states~\cite{Cheng2016} and the collective mode associated with the phase rather than the amplitude~\cite{Seibold2017,Feigelman2018}. 

Recently, it has been theoretically shown that under the injection of dc supercurrent, the Higgs mode linearly interacts with ac electric field polarized along the direction of the supercurrent flow. 
Accordingly, the Higgs mode is predicted to appear in the linear response function such as the optical conductivity spectrum $\sigma_1(\omega)$, giving rise to a polarization-dependent peak structure at the superconducting gap frequency $\omega=2\Delta$~\cite{Moor2017}.  
This effect comes from the momentum term in the action:  
\begin{equation}
S=\int C\mathbf{Q}^2 (t) |\Delta(t)|^2 dtd\mathbf{r} 
\end{equation}
where $\mathbf{Q}(t)=\mathbf{Q}_0+\mathbf{Q}_\Omega(t)$ is the gauge-invariant momentum of the condensate consisting of the dc supercurrent term $\mathbf{Q}_0$ and the ac electric-field (probe-field) driven term $\mathbf{Q}_\Omega(t)=\text{Re}[\mathbf{Q}_\Omega \exp(i{\Omega}t)]$, respectively, and 
$\Delta(t)=\Delta_0+\delta\Delta(t)$ is the time-dependent superconducting order parameter. 
The action $S$ includes the integral of 
$\delta\Delta_{2\Omega}\mathbf{Q}^2_{-\Omega}$ 
and 
$\delta\Delta_{\Omega}\mathbf{Q}_{0}\mathbf{Q}_{-\Omega}$ where $\delta\Delta_{\Omega(2\Omega)}$ 
denotes the Fourier component of the oscillating order parameter (Higgs mode). 
The first term corresponds to the quadratic coupling of the Higgs mode to the gauge field which was previously demonstrated~\cite{Matsunaga2014}. 
The second term indicates linear coupling between the Higgs mode and the gauge field induced by the finite amount of condensate momentum $\mathbf{Q}_0$ parallel to the probe electric field. 
It is then expected that the polarization-selective excitation and detection of the Higgs mode are attainable within the linear response regime when supercurrent is injected into the system. 
Motivated by this theoretical prediction, here we investigated the effect of supercurrent injection on the optical conductivity in an s-wave superconductor, NbN thin film.

\begin{figure}[bp]
\includegraphics[width=.49\textwidth]{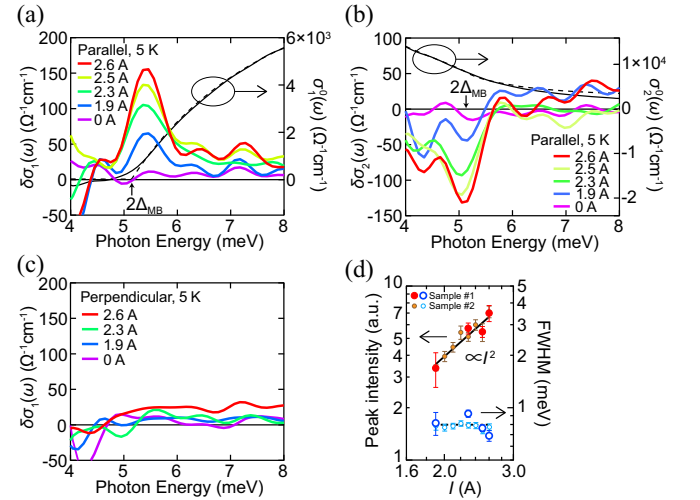}%
\caption{The change of the (a) real part and (b) imaginary part of the optical conductivity spectra induced by supercurrent injection taken with the THz probe electric field polarized along the direction of supercurrent. 
The optical conductivity spectra measured without the current are also plotted in (a) and (b) with the right axes. 
The dashed line depicts the Mattis-Bardeen fit and the superconducting gap estimated from the fit is marked by the vertical arrow. 
Shown in (c) is the change of the real part of the optical conductivity spectra taken with the THz probe electric field perpendicular to the current direction.  
Current ($I$) dependences of the peak intensity and FWHM ($=2\sqrt{2\log2}\sigma)$ are plotted in (d). 
The lines are trend lines determined by the method of least squares. 
Details are found in the text.
Small points in (d) are data of another sample (Sample \#2) fabricated in very similar conditions~\cite{SI}.}\label{fig2}
\end{figure}

The optical conductivity was measured using terahertz time-domain spectroscopy (THz-TDS) technique in transmission geometry. 
The schematic diagram of the experimental setup is shown in Fig~\ref{fig1}(b). 
The sample is an epitaxial NbN film of 26\,nm in thickness grown on a 400-$\mu$m-thick MgO (100) substrate. 
The critical temperature is 14.5$\pm$0.2\,K as confirmed by the transport and magnetic susceptibility profiles (see Fig.~\ref{fig1}(c)). 
The transition width $\Delta T_c/T_c$ is only about one percent estimated from the magnetic susceptibility, indicating the high uniformity of the film. 
The supercurrent was injected through the Au/Ti electrodes deposited on both ends of the sample 
and the critical current is 3.8\,A at $T=$5.1\,K (3.7\,MA/cm$^2$ in current density). 
The sample was cooled down using a $^4$He flow cryostat in a $^4$He atmosphere.  
For the measurements, the laser beam from the mode-locked Ti:sapphire oscillator (repetition rate: 80\,MHz, wavelength: 800\,nm, average power: 1\,W, pulse width: 100\,fs) was split into two with the intensity ratio of 3:1; one for the THz generation modulated by an optical chopper rotating at $\approx$2.3\,kHz and one for the gate pulse for the electro-optic (EO) sampling. 
The probe THz pulse was generated by optical rectification in a ZnTe crystal, and detected by EO sampling also using a ZnTe crystal. 
The peak value of the probe THz electric field was below 20\,V/cm which is weak enough to assure the linear response regime. 
Polarization of the THz pulse is determined by wire grid polarizers placed before and after the sample. 
The thin solid line in Fig.~\ref{fig2}(a) represents the optical conductivity $\sigma_1^0(\omega)$ measured at 5\,K without current injection. 
The superconducting gap structure is clearly identified at $\approx5$\,meV. 
The spectrum is well fitted by the Mattis-Bardeen model with the gap energy of 2$\Delta_{\text{MB}}=5.14\pm0.01$\,meV, as shown by the dashed line in Fig.~\ref{fig2}(a)~\cite{Zimmermann1991}.

To measure precisely the conductivity change $\delta\sigma_1(\omega)$ induced by the current injection with eliminating the long-term fluctuation, we repeated the THz waveform scan with and without the current injection alternatingly, and accumulated the waveform from 30--100 times. 
The differential spectrum $\delta\sigma_1(\omega)$ is extracted from the Fourier transform of the waveforms with and without the current, $E^{\text{I}}(\omega)$ and $E^0(\omega)$, respectively, the refractive index of the substrate $n_{\text{sub}}(\omega)$, and the premeasured $\sigma_1^0(\omega)$ without current injection, using the following equation commonly used for pump-probe measurements for thin film samples~\cite{Matsunaga2012}: 
\begin{equation}
{\delta\sigma}_1 ({\omega})=\frac{1+n_{\text{sub}} ({\omega})+Z_0 d{\sigma}_1^0 ({\omega})}{Z_0 d} \left(\frac{E^0(\omega)}{E^{\text{I}}(\omega) }-1\right),
\end{equation}
where $Z_0=377\,\Omega$ is the vacuum impedance, and $d$ is the thickness of the NbN film. 
At low temperature the transmission is very low below the gap energy so that the data below 4\,meV are scattered~\cite{SI}. Therefore in the following graphs the data above 4\,meV are plotted. 

The thick lines in Fig.~\ref{fig2}(a) show the differential spectra of $\delta\sigma_1(\omega)$ under the currents $I=$0, 1.9, 2.3, 2.5, and 2.6\,A at 5\,K ($=0.34T_c$) measured for the polarization parallel to the direction of the supercurrent\footnote{The oscillating structure appearing in all the differential spectra is attributed to interference effects of the THz-TDS measurements exaggerated by the differentiation. This already exists in the equilibrium spectra.}. 
A peak structure is clearly identified in all the data with currents between 1.9 and 2.6\,A. 
Note that, 2.6\,A is 70\% of the critical current $I_c$ at the temperature. 
The peak height is at most 1\% with respect to the normal state conductivity $\sigma_N$=1.2$\times$10$^4$\,$\Omega^{-1}$cm$^{-1}$. 
The peak position is estimated as 5.40$\pm$0.05\,meV (1.30\,THz), which is slightly larger than the onset energy of $\sigma_1^0 (\omega)$ by amount of 0.2\,meV, while it coincides with the energy of Higgs mode estimated from the time-resolved observation of the Higgs mode oscillation in previous pump-probe measurements in another NbN film with similar thickness (24\,nm)~\cite{Matsunaga2013}. 
Here we take into account the effect of thermal broadening~\cite{Murakami2016Damping} practically by convoluting a Gaussian distribution to the original Mattis-Bardeen spectrum function. 
The broadening width is estimated as 0.6\,meV at 5\,K and 2\,meV at 13\,K, which can explain the slight energy difference between the peak center and the $2\Delta_{\text{MB}}$. 
Although the supercurrent-induced conductivity peak shows a tail in the higher energy side like the theoretically predicted one~\cite{Moor2017}, here we fitted the peak assuming a Gaussian function with a constant offset for each current density: $a\cdot g(x)+b$ with $g(x)=1/\sqrt{2\pi\sigma^2} \exp(-((x-\mu)/\sigma)^2/2)$. 
While the peak center $\mu$ and width $\sigma$ are almost constant, the peak intensity $a$ increases with the current as shown in Fig.~\ref{fig2}(d). 
The solid line indicates the quadratic fit to the peak intensity and the dashed line indicates the average of the peak width determined by the method of least squares. 
It should be remarked that the peak width, 0.80$\pm$0.05\,meV, does not vary with the temperature up to 8\,K (not shown), suggesting the peak width is hardly affected by the thermal broadening effects mentioned above. 
Notably, this peak structure is completely absent when measured for the perpendicular polarization at the same temperature and currents as indicated in Fig.~\ref{fig2}(c). 
These characteristics are consistent with the theoretical prediction that the spectral weight of the current-induced Higgs mode resonance should be proportional to $I^2\cos^2\theta$ where $\theta$ is the angle between the supercurrent flow and the THz electric field~\cite{Moor2017}. 

To establish more firm connection between the theory and experiment, we calculated the complex optical conductivity based on the theoretical work by Moor~\textit{et~al}.~\cite{Moor2017} with the parameters relevant to the present experiments at $T=$5\,K. 
To simulate realistic optical conductivity spectra, we introduced a broadening factor $\Gamma$ for the peak by convoluting the ideal conductivity spectra with Gaussian distribution of $2\Delta$ with FWHM ($=\Gamma$) of 0.6\,meV. 
According to Ref.~\citenum{Moor2017}, the peak weight is given by a coefficient $W_Q=D Q_0^2$ with the diffusion constant $D=4k_B/(\pi e) \left|dH_{c2}/dT\right|^{-1}=9.0\times10^{-5}$\,m$^2$/s determined from $H_{c2}$ measured~\cite{Semenov2009}. 
The condensate momentum $Q_0$ is defined as $m v_s/\hbar$ where $v_s$ is the condensate velocity calculated from the superfluid density $n_s$ and the injected current density $J$ using $J=e n_s v_s$. 
The $n_s$ is estimated as $5.4\times10^{26}$\,m$^{-3}$ from the imaginary part of the optical conductivity spectrum $\sigma_2(\omega)$ using $n_s=\displaystyle\lim_{\omega\to0}m\sigma_2 (\omega)\omega/e^2$. 
When the current $I$ is $2.6$\,A, the $W_Q$ is estimated as $2.3\times10^{-3}$\,meV, which is as small as $4.3\times10^{-4}\Delta$ at $T=5$\,K so that the theory in Ref.~\citenum{Moor2017} is applicable. 

\begin{figure}[bt]
\includegraphics[width=.49\textwidth]{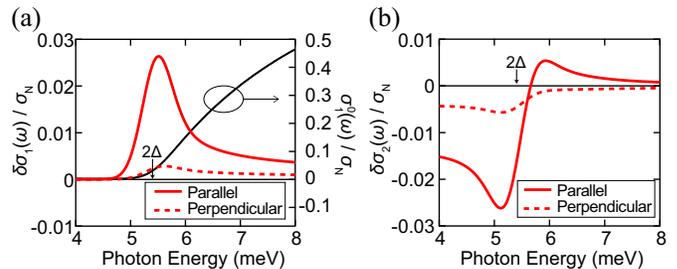}%
\caption{Theoretically expected changes of (a) real and (b) imaginary parts of the optical conductivity induced by supercurrent injection with respect to the normal state conductivity $\sigma_N$~\cite{Moor2017}. 
We set $W_Q = 0.00043\Delta$ (corresponding to 2.6\,A), $T =5$\,K, the mean of $2\Delta=5.4$\,meV, and $\Gamma=0.6$\,meV. 
The optical conductivity without current injection is also represented as a thin line in (a).}\label{fig3}
\end{figure}

Figures~\ref{fig3}(a) and (b) show the real and imaginary parts of the calculated spectra, respectively. 
Compared to the Figs.~\ref{fig2}(a) and (b), the calculation reasonably reproduces the characteristic peak in $\delta\sigma_1$, dispersive shape in $\delta\sigma_2$, and gradual onset of $\sigma_1^0$ at a slightly below 2$\Delta$. 
The calculation also indicates that the spectral weight is transferred from the condensate at zero energy to the $2\Delta$ peak (not shown in the figure). 
The transferred spectral weight is only a few percent of that of the whole condensate. 
The calculation shows about one order smaller signal for the perpendicular configuration (dashed lines in Figs.~\ref{fig3}(a) and (b)), which is unrecognizable in our experiments because of the detection limit. 
The calculation is in reasonable agreement with the experimental result even quantitatively, suggesting that the observed peak is attributed to that of Higgs mode.

Now we address other possible mechanisms that could also give a conductivity peak structure.   
Firstly, the phase mode is not plausible as it is lifted to the plasma frequency at low temperature limit due to the Anderson-Higgs mechanism. 
In fact, it can be lowered to the energy region near the gap, known as the Carlson-Goldman mode~\cite{Carlson1975}, but only at temperature close to $T_c$, which is not relevant to our results taken substantially below $T_c$. 
Single particle excitations caused by the current injection can also give rise to the change of optical conductivity particularly near the critical current density. 
They cause smearing of the density of states and shrinkage of the gap which can induce a singularity in the optical conductivity spectrum at around the gap energy~\cite{Semenov2016}. 
Indeed in the previous experiments on impure aluminum, an absorption peak was observed at slightly lower energy than the pristine gap under the presence of strong in-plane magnetic fields ($H>0.5H_c$) which induced effective in-plane supercurrents~\cite{Budzinski1966,Budzinski1973}. 
Under such a strong field, the conductivity peak is explained within the framework of quasiparticle excitations~\cite{Ovchinnikov1978}.   
On the other hand, in our NbN, the maximum current density corresponds to a very weak magnetic field of $0.02H_{c1}=3\times10^{-5}H_{c2}$, therefore, as suggested by Moor et al.~\cite{Moor2017} the Higgs mode response dominates the observed conductivity peak. 

\begin{figure}[tb] 
\includegraphics[width=.49\textwidth]{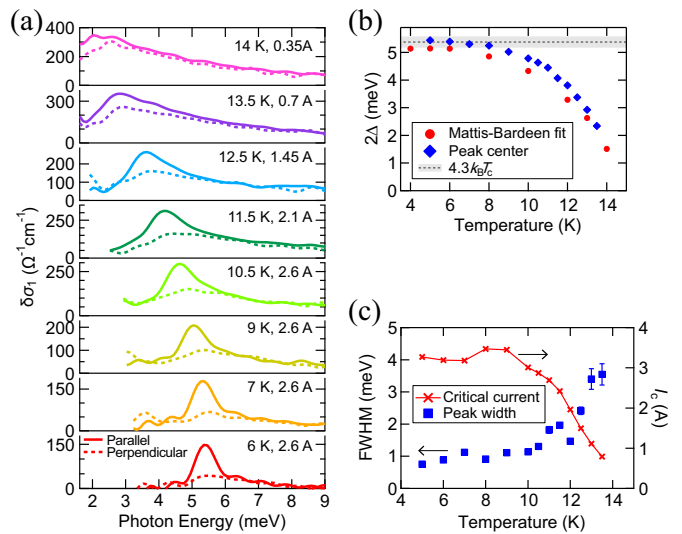}%
\caption{(a) Change of the real part of the conductivity $\delta\sigma_1(\omega)$ induced by the supercurrent injection with the THz probe electric field parallel (solid lines) and perpendicular (dashed lines) to the current direction at the indicated temperatures and currents. 
(b) Temperature dependence of the peak center and the superconducting gap $2\Delta_{\text{MB}}$ determined from the Mattis-Bardeen fit to $\sigma_1^0(\omega)$~\cite{Zimmermann1991}. The horizontal line shows the zero-temperature superconducting gap estimated from the $T_c$.
(c) Temperature dependence of the peak width (square; left axis) and the critical current $I_c$ (cross; right axis).}\label{fig4}
\end{figure}

Finally, we extended the measurement up to 14\,K ($=0.96T_c$). 
The real part of the conductivity change $\delta\sigma_1(\omega)$ is plotted in Fig~\ref{fig4}(a). 
With increasing the temperature,  
the energy of the peak center gradually decreases in synchronization with the superconducting gap $2\Delta_{\text{MB}}$ determined from the Mattis-Bardeen fit to $\sigma^0(\omega)$ as shown in Fig~\ref{fig4}(b). 
The figure also shows the zero-temperature gap $2\Delta_0$ estimated as  $4.3k_{\text{B}}T_{c}$~\cite{Chockalingam2009} plotted as a horizontal line, which agrees very well with the energy of the peak center at low temperature. 
The peak width is almost constant at low temperature ($T\le10$\,K), then rapidly increases at higher temperature.
This temperature dependence shows a negative correlation with that of the critical current as shown in Fig.~\ref{fig4}(c), presumably because both the Higgs mode and superconducting current are affected by thermally-induced quasiparticles. 
The polarization dependence becomes less significant at higher temperature, which is qualitatively consistent with the theoretical expectation~\cite{SI}.
Note that the sample temperature is precisely controlled during the measurements to negate Joule heating effects because the differential signal $\delta\sigma_1(\omega)$ is very sensitive to the sample temperature at $T>8$\,K where the temperature coefficient of the gap has a non-zero value~\cite{SI}.

In summary, we have demonstrated that the Higgs mode appears in the linear optical conductivity spectrum under the supercurrent injection. 
A distinct peak slightly above the optical conductivity gap accompanied by a high energy tail is clearly observed in the film of an s-wave superconductor NbN. 
Based on the polarization and current density dependencies, we attribute the peak structure to the Higgs mode as recently suggested in the theoretical study~\cite{Moor2017}. 
The peak energy is also in agreement with the oscillation frequency of Higgs mode observed in previous time-resolved THz pump-THz probe measurements~\cite{Matsunaga2013}. 
This method comprising the linear conductivity measurement and the current injection provides a new pathway to access the Higgs mode in various superconductors.
Extension of the measurement scheme to p- or d-wave superconductors would be highly intriguing. 
The method may also be applied to study the superconductivity with competing orders in unconventional superconductors.

\begin{acknowledgments}
This work was supported in part by JSPS KAKENHI (Grants Nos. 15H02102, 15H05452, 18K13496), and by the Photon Frontier Network Program from MEXT, Japan. 
Y. M. is supported by JSPS Research Fellowship for Young Scientists.
\end{acknowledgments}


\end{document}